\documentstyle[prl,aps]{revtex}

\begin{document}
\title{New Monte Carlo Iteration Method for Generalized Ensembles}

\author{Jesper Borg}
\address{Niels Bohr Institute Blegdamsvej 17, DK-2100 {\O}, Denmark; \\ Condensed Matter Physics and Chemistry Department, Ris\o\ National Laboratory, Denmark.}
\maketitle

\begin{abstract}

We present a new and general Monte Carlo iteration method for generalized ensembles. 
It consists of two elements: (1) a simple algorithm to distinguish between distributions arising 
from respectively equilibrium- and non-equilibrium processes, and 
(2) a selfconsistent maximum-likelyhood estimation of the unknown 
thermodynamic quantities, based  on the information obtained at all previous iterations. 
We demonstrate the efficiency of the method by calculating the density of state function 
of a homopolymer with $16^3$ monomers. This represents an
improvement of at least an order of magnitude compared to previous studies.
\vspace{0.5cm}


\end{abstract}

\vskip0.5cm
%--------------------------------------------------------------------

The method of Monte Carlo (MC) integration has proven 
succesfull for calculating thermodynamic properties of model systems
with moderate number of degrees of freedom \cite{Hesselbo}.
The basic idea is to sample the phase space by generating a Markov chain
of states through a fixed matrix of transition probabilities. 
These probabilities are chosen
such that the condition of detailed balance is fulfilled for the statistical ensemble in concern.  
For instance, the traditional Metropolis algorithm \cite{metropolis} samples 
directly in the canonical ensemble (CE) by the choise of Boltzmann transition 
probabilites. 

A generic deficiency of the Metropolis sampling technique is 
the {\it slow relaxation} of the Markov chain which typically appears 
at points of phase transitions or at low temperatures. For systems  
with a rugged energy landscape, such as spin glasses \cite{parisi}, 
random heteropolymers \cite{shak} or even spin-interacting homopolymers \cite{borg2001}, 
this problem effectively causes the chain to be {\it trapped} in the phase space around 
local energy minima. In any case, slow relaxation easily leads to results which are 
errorously sensitive to the initial states of the Markov chain.

In the past decade, a variety of MC-methods, based on non-Boltzmann probability distributions, 
have been developed to improve the phase space sampling,
They are commonly referred to as broad energy ensemble or {\it generalized ensemble} (GE) methods 
\cite{hansmann97} corresponding to the generalization of the partition function, $Z$: 
\begin{equation}
\label{genZ}
Z=\sum_\phi \Gamma(\phi)\omega(\phi)~.
\end{equation}
Here, $\phi$ is some coarse-grained state variable, $\Gamma$ is the density of
states and $\omega(\phi)$ are the ensemble-defining weights.
Different choices have been proposed in the litterature.
In the multicanonical approach (MU) \cite{berg}, 
$\phi$ is simply the energy, $E$, of the system and
$\omega_{MU}(E)=\Gamma(E)^{-1}$. This approach also 
includes the entropic sampling \cite{Lee,berg2}. 
In the so-called $1/k$-ensemble \cite{Hesselbo} the weights are defined 
as $\omega_{1/k}(E)=k(E)^{-1}$, with
$k(E)=\sum_{E'\le E}\Gamma(E')$. Finally, for 
Simulated Tempering (ST) \cite{Lyobartsev,Marinari}  $\phi$ also
contains a temperature index $l$, i.e. $\phi=\{E,l\}$, and 
$w_{ST}(E,l)=\exp(-\beta_l E+f_l)$, where $f_l$ is the reduced free 
energy at the inverse temperature $\beta_l$,
$$
e^{-f_l}=Z_{CE}(\beta_l)= \sum_E \Gamma(E) e^{-\beta_l E}~.
$$
Here, $Z_{CE}(\beta)$ is the usual canonical partition function.
By inspection of eq. (\ref{genZ}) it is clear that MU is equivalent to
uniform energy sampling and ST is equivalent to a uniform temperature sampling
in the preassigned set of values $\{ 1/\beta_l\}_l$. 
The last method practically corresponds to a uniform entropy sampling
\cite{Hesselbo}. 

Unlike the canonical ensemble, the probability weights are not a
priori known in the generalized ensembles, since $\omega$ in all cases is a
function, $g$, of the density of states (or the free energy), $\omega=g(\Gamma)$, 
and the knowledge of $\Gamma$ corresponds exactly to solving the thermodynamics. 
The standard solution to this problem is to determine $\omega$ through an iterative procedure, 
$\omega_{i+1}=F_1(N_i,\omega_i)$, where $N_i=N_i(\phi)$ is the observed histogram over $\phi$ 
obtained in the $i$'th simulation with the weights $\omega_i$. The map, $F_1$, is contructed to 
have the desired fixed point, $F_1(\bar{N},g(\Gamma))=g(\Gamma)$, where $\bar{N}=\bar{N}(\phi)$ 
is the expected number of counts in a simulation with weights $g(\Gamma)$. 
Typically, the iteration process is stopped when the observed
histogram and the weights are sufficiently close to the fixed point requirement.
 A large run is then performed in which the physical quantities of interest are studied.

Despite several succesfull applications of this iteration scheme 
(see eg. \cite{Hesselbo,hansmann97,berg,Lee,berg2,Lyobartsev,Marinari,Irback99}), 
it suffers from the {\it loss of statistics} inhereted in the updating rule for the weights.
Since $\omega_{i+1}$ only depends on the last histogram, $N_i$, a nescessary requirement to 
insure convergence is that the set ${\cal S}(N_i)=\{\phi | N_i(\phi)>0\}$ encloses at least 
the same region of the (coarse-grained) phase space as the previously obtained set ${\cal S}(N_{i-1})$, 
i.e. ${\cal S}(N_i)\supseteq {\cal S}(N_{i-1})$. Otherwise, the iteration $F_1(N_i,\omega_i)$ will 
lead to a worse estimate of $g(\Gamma)$ for such regions of the phase space.  
This is clearly a strong statistical requirement. In particular, it prevents for a more adiabatic 
updating of the weights, which otherwise could speed up the convergence. 

In this letter we propose a new iteration scheme which accounts for the information 
obtained from in principle {\it all} previous iterations. More precisely,  a selfconsistent map 
$\omega_{i+1}=F_M(\{ N_{(i-j)},\omega_{(i-j)} \}_{j=0}^{M-1})$, will be derived where the 'memory' $M$ 
can be chosen arbitrary large.
As such there will not be any specific statistical requirement to insure the  
convergence, and it therefore allows for an extension of the GE-analysis to systems of much 
greater complexity or size.

The scheme is based on a maximum-likelyhood estimate of the density of state consistent 
with the set of distributions $\{N_i\}_i$. 
In order not to overestimate metastable states, it is important, however, to distinguish between 
statistics arising from respectively equilibrium and non-equilibrium processes. Therefore we will 
first discuss a simple but general algorithm for doing so. Its basic principle is to assign a number 
$\theta=\theta(\phi)$ and a set $\Theta(\phi)$ to each $\phi$-bin during the simulation.
This pair, $(\theta,\Theta)$, is updated on the same MC-rate as the distribution $N_i$, say 
after each timestep $\Delta t$. If $\phi_t$ denotes the state of the system at time $t$ and 
$\phi'_t$ is the test state for the last move (in case of acceptance, $\phi'_t=\phi_t$) the 
quantities $(\theta,\Theta)$ are given by the following rules:

\begin{itemize}

\item[(1)]{Initially, $(\theta,\Theta)(\phi)\stackrel{def}{=}(0,\{\})$ for all states $\phi$ except 
for the first one $\phi_0$, for which $(\theta,\Theta)(\phi_0)\stackrel{def}{=}(1,\{1\})$.} 
\item[(2)]{If the transition $\phi_t\rightarrow \phi_{t+\Delta t}$ takes the Markov chain into a 
new state, $\theta(\phi_{t+\Delta t})=0$, one puts 

$(\theta,\Theta)(\phi_{t+\Delta t})\stackrel{def}{=}
(\theta(\phi_t)+1,\Theta(\phi_{t+\Delta t})\cup\{\theta(\phi_t)+1\})$.}

\item[(3)]{If the transition $\phi_t\rightarrow \phi_{t+\Delta t}$ takes the Markov chain into an 
old state, $\theta(\phi_{t+\Delta t})>0$, one redefines $(\theta,\Theta)$ for all states $\phi$ 
for which $\theta(\phi)>\theta(\phi_{t+\Delta t})$ as 
$\theta(\phi)\stackrel{def}{=}\theta(\phi_{t+\Delta t})$ and
$\Theta(\phi)\stackrel{def}{=} \{\theta(\phi_{t+\Delta t})\}\cup 
\{ m \in \Theta(\phi)~|~ m \le \theta(\phi_{t+\Delta t})\}$.}

\item[(4)]{After the value $\theta(\phi_{t+\Delta t})$ has been determined by (2) or (3) 
$\Theta(\phi'_{t+\Delta t})$ is seperately updated as 

$\Theta(\phi'_{t+\Delta t})\stackrel{def}{=}\Theta(\phi'_{t+\Delta t})\cup \{\theta(\phi_{t+\Delta t})\}$.}

\end{itemize}

It follows, that the $\theta$-value of the present state $\phi_t$ of the simulation always 
equals the maximum value. The purpose of the $\theta$-function is to partition the phase space 
into the {\it kinetically connected} or locally equilibrated regions. Each set 
$\tilde{\Phi}_m=\{\phi|\theta(\phi)=m\}$ defines such a region, whereas the transition 
$\tilde{\Phi}_m\rightarrow \tilde{\Phi}_{m'}$ with $m'>m$, corresponds to a process which, 
within the total simulation time $\tau_i$ spent at iteration $i$, effectively is out of equilibrium. 
The time-step $\Delta t$ for this analysis should be large enough to reflect the local 
properties of the free energy landscape, $f(\phi)\sim \log(\Gamma(\phi)\omega(\phi))$, 
i.e. $\Delta t$ should be comparable to the (local) decorrelation time. 
Dependent on the choise of $\Delta t$, some of the regions generated by the values of 
$\theta$ might contain only very few counts. This will for instance be characteristical 
for the transition region between two minima of the free energy landscape. 
After the completion of the $i$'th simulation, a simple minimums criteria on the statistical content 
can be applied \cite{borg1} to discard such values of $\theta$. Hereafter, $\theta$ can be considered 
as a way of labelling the different effective 'macro-states' or basins \cite{sneppen,berry} 
observed in the simulation.  

The 'adjoint' quantity, $\Theta$, keeps track of the neighbourhood of each basin as well, 
$\Phi_m=\{\phi~|~m\in\Theta(\phi)\}\supseteq \tilde{\Phi}_m$, by the inclusion of the observed
 but possibly rejected states (rule 4). The nescessity for this rather tedious book-keeping is 
that the rejection of a state may indicate that the corresponding free energy previously has been 
underestimated. The algorithm is schematically illustrated in fig. 1. 

Following the line of arguments, the distribution $N_i(\phi)$ resulting from iteration $i$ 
must be partitioned, $N_i\rightarrow \{N_{im}\}_m$, according to the different equilibrium regions 
$\{\tilde{\Phi}_m \}_m$. If the simulation is done in the ST-ensemble, a natural further partioning, 
$m\rightarrow m'=ml$, is defined by the different temperatures, $T_l$; such that $\{\Phi_{im'}\}_{m'}$ 
and $\{N_{im'}\}_{m'}$ can be considered as sets and functions in energy space alone. 
For all types of ensembles, let $M_i$ denote the total number of partitions at iteration $i$ and 
let $m_i$ be the corresponding index variable. Furthermore, put $s=\{i,m_i\}$ such that a particular 
set or histogram obtained at a given iteration can be referred to simply as $\Phi_s$ or $N_s$ respectively.

The map, $F_M$, is readily constructed in the following way. 
The probability $p_s(E~|\Gamma)$ of observing the energy $E$
within the domain $\Phi_s$, given the density of states $\Gamma$ 
and the weights $\omega_s(E)$, is defined by

\begin{eqnarray}
p_s(E | \Gamma)=\frac{\Gamma(E)\omega_s(E)}{Z_s},~~~~
Z_s\stackrel{def}{=}\sum_{E\in \Phi_s} \omega_s(E)\Gamma(E)~. \nonumber 
\end{eqnarray}

Under the assumption of (restricted) statistical independence, the histogram $N_s(E)$ will be a member 
of the multinomial \cite{adv} probability distribution $P_s$;
\begin{eqnarray}
P_s(N_s| \Gamma)=n_s ! \prod_{E \in\Phi_s} \frac{p_s(E|\Gamma)^{N_s(E)}}{N_s(E)!}~,\nonumber
\end{eqnarray}
where $n_s=\sum_E N_s(E)$ is the total number of counts in $\Phi_s$. 
The {\it likelyhood} ${\cal L}$ for observing  the set of histograms $\{N_s\}_{s=1}^M$  ($M=\sum_i M_i$), is given by the product of the $P_s$'s ,
i.e. ${\cal L}(\{N_s\}_{s=1}^M | \Gamma )=\prod_{s=1}^M P_s(N_s| \Gamma)$. 
An efficient \cite{adv} estimate,  $\hat{\Gamma}$, of the true density of states can now be obtained 
by maximizing ${\cal L}$ with respect to $\Gamma$ . This leads to the expression,
\begin{equation}
\label{gamma}
\hat{\Gamma}(E)=\frac{\sum_{s=1}^M N_s(E)}
{\sum_{s=1}^M \delta(E\in \Phi_s)n_s\omega_s(E)Z_s^{-1}}~,
\end{equation}
where $\delta(E\in \Phi)=1$ if $E\in\Phi$ and zero otherwise.
The partition functions $Z_s$ must be estimated selfconsistently from eq. (\ref{gamma}). 
This set of estimates $\{\hat{Z}_s\}$ can be expressed as the solution to the following $M$ equations, 
$y_{s=1,\ldots,M}$;
\begin{equation}
\label{multi}
y_s\stackrel{def}{=}\sum_{E\in \Phi_s} \frac{\sum_{t=1}^M N_t(E)}
{\sum_{t=1}^M \delta(E\in \Phi_t)n_t\frac{\omega_t(E)}{\omega_s(E)}\frac{\hat{Z_s}}{\hat{Z_t}}}=1~.
\end{equation}
In fact, these equations contain the {\it multihistogram equations} \cite{ferrenberg89} as a special case.
 This comparison together with the different numerical solving procedures mentioned below, will be
 discussed in details in a seperate paper \cite{borg1}.

By inspection of eq. (\ref{multi}) it is clear that each disconnected region of 
$\Phi_{tot}=\bigcup_{s=1}^M \Phi_s$  will introduce a zero mode in the Jacobian 
$\frac{\partial y}{\partial Z}$, corresponding to a (local) normalization constant in terms of an 
entropy or free energy, which can be chosen freely. In particular, an overall normalization constant 
has to be fixed in order to define a unique solution. A way around the problem of multiple zero modes, 
is to solve eq. (\ref{multi}) seperately for each connected region and estimate the relations between 
the normalization constants afterwards by interpolating the slope of the entropy function 
$\hat{S}(E)=\log(\hat{\Gamma}(E))$. Obviously, a reliable estimate of $\Gamma$ is not obtained before 
$\Phi_{tot}$ constitutes one overall connected set. 

The weights for simulation $i+1$ is obtained by inserting the solution 
$\{\hat Z_s\}^M_{s=1}$ ($M=\sum_{i'=1}^i M_{i'}$) of eq. (\ref{multi}) into expression (\ref{gamma}) 
and use $\omega_{i+1}=g(\hat{\Gamma})$. Contrary to the standard GE-iteration method, 
the Markov chain does not have to be reinitialized for the next iteration, not even if the Markov chain 
at iteration $i$ ends in a trapped region of the phase space. Since the iteration scheme for $\omega$ 
keeps all the information previously obtained and since the new weights $\omega_{i+1}$ by construction 
improves the phase space sampling, one can simply continue the next iteration from the final state 
$\phi_{\tau_i}$ of the last one. This makes the scheme very efficient in optimization problems. 
For the sake of completeness it should also be noted, that both the 'equilibrium sorting' algorithm 
and eq. (\ref{gamma})-(\ref{multi}) can be directly applied to multidimensional state variables 
$\vec{\phi}$ as well.

As a demonstration of the method we calculate the density of states of a self-avoiding homopolymer 
with a nearest-neighbour (non-bonded) potential 
of energy $\epsilon=-1$ in a simple cubic lattice \cite{flory}. Homopolymer models of this type have 
recently been the subject of extensive studies, both on-lattice \cite{meirovitch,yuan91,lai,grassberger}
 and off-lattice \cite{Irback99,grassberger95,zhou}. Order parameter related quantities have been 
calculated in \cite{grassberger} up to rather long on-lattice chains ($L=5000$), 
but estimates of heat-capacities or densities of states have been limited to much shorter chains 
(in all cases $L<400$), due to the normal difficulty associated with obtaining converged results 
for these quantities \cite{frantz}. 

In this study, we have chosen $L=16^3$ and performed three independent runs in the 
multicanonical ensemble, $g(\Gamma)=\Gamma^{-1}$, with the parameters 
$\tau_{tot}=\sum_i \tau_i\simeq 4\cdot 10^9$ and $\Delta t=4\cdot 10^3$. 
The energies are binned with $\Delta E=8$ which implies that $\sim 8\cdot 10^2$ independent values of 
$\Gamma$ has to be estimated. For each run the system is initialized by a random self-avoiding walk and 
with  the infinite temperature weights $\omega_{i=1}(E)=1$. The simulation time $\tau_i$ in each iteration 
within a run is defined dynamically, roughly proportional to the size of the phase space observed 
\cite{borg1}.  The maximal 'memory' of the iteration scheme is set to $M_{max}=100$, which has turned out 
to be sufficient to contain the full history of each run. 

The results are shown in fig. 2. In the large figure the change of entropy $\Delta S$ compared to the 
random coil is plotted as function of the total energy $E$, for the three runs.  Deviations can be 
observed for energies $E<-5500$, but otherwise the differences are less than the size of the data points. 
In the inset, these differences are depicted in terms of $~\Delta\Delta S=\Delta S-(\Delta S)_{av}$, 
where $(\Delta S)_{av}$ is the weighted average of the three runs. For energies $E>-5500$ 
the relative uncertainties are generally observed to be of the order $\sim 10^{-3}$. 
No thermodynamical transitions takes place below $E=-5500$, and the results have converged in 
the sense that they are sufficiently accurate to reproduce the sensitive thermodynamic quantities
 such as the temperature dependence of the heat capacity. We refer to a forthcoming publication 
\cite{borg2} for this analysis. Keeping in mind the level of convergence and relating the size of 
the system to the total simulation time, the iteration method constitutes an improvement of at
 least one order of magnitude compared to previous studies.
  
In fig. 3, a typical state of the homopolymer in the energy region $E<-5500$ is depicted. 
The state is completely compact, and its difference to the ground state (the box $16^3$) is only due 
to the particular arrangement of the 'rigid' surface. The structure of the state testifies to the fact 
that the method is capable of investigating most of the coarse grained phase space, even within the 
relatively short MC-time $\tau_{tot}$ available. 
 
In summary, we have presented a new and general Monte Carlo iteration scheme for generalized ensembles 
(GE), which allows for an extension of the GE-analysis to systems of much greater size or complexity. 
In the future, we plan to test the method on off-lattice polymers and spin-glasses.

It is a pleasure for me to acknowledge the helpful discussions with M.H. Jensen, K. Sneppen, P.-A. Lindgaard, G. Tiana and A.D. Jackson.

\newpage

\begin{figure}
\caption{Schematic illustration of the equilibrium sorting algorithm.
({\bf a}) shows the free energy landscape $f(\phi)$ (in arbitrary units) with two minima as function 
of the state variable $\phi$ (arbitrary numbers) for some system, which is initialized in state $\phi_0$.
({\bf b}) shows a possible time evolution of the system in a MC-simulation. 
The full line is the actual history, whereas the marks (+) show the test states of the Markov chain. 
Two non-equilibrium transitions are observed, $\phi_0 \rightarrow \tilde{\Phi}_1$ and 
$\tilde{\Phi}_1 \rightarrow \tilde{\Phi_2}$. Here, $\tilde{\Phi}_1$ and $\tilde{\Phi}_2$ are the basins 
defined by the statistical relevant values of $\theta$ (see text). The figure also shows the extended 
regions $\Phi_1$ and $\Phi_2$ obtained by the inclusion of the test states within each basin.}
\end{figure} 

\begin{figure}
\caption{
The change of entropy $\Delta S$ compared to the random coil for a homopolymer with $16^3$ monomers, 
as function of the total energy $E$. Three independent simulations have been performed with 
$\sim 4\cdot 10^9$ MC-steps. In the inset the differences between the results are shown in therms of 
$~\Delta\Delta S=\Delta S-(\Delta S)_{av}$, where $(\Delta S)_{av}$ is the weighted average of the 
three runs. The relative uncertainty of $\Delta S$ is generally observed to be of the order
 $\sim 10^{-3}$ for energies $E>-5500$.}
\end{figure}

\begin{figure}
\caption{
An example of a typical homopolymer state with $L=16^3$ monomers
in the low energy region $(E<-5500)$. Only the particular arrangement of the 'rigid' surface makes 
it different from the true ground state. The structure of the state testifies to the fact, 
that the simulation is capable of investigating most of the relevant phase space within a relatively 
short MC-time $\sim 4\cdot 10^9$.}
\end{figure}

% --------------------------------------------------------------------

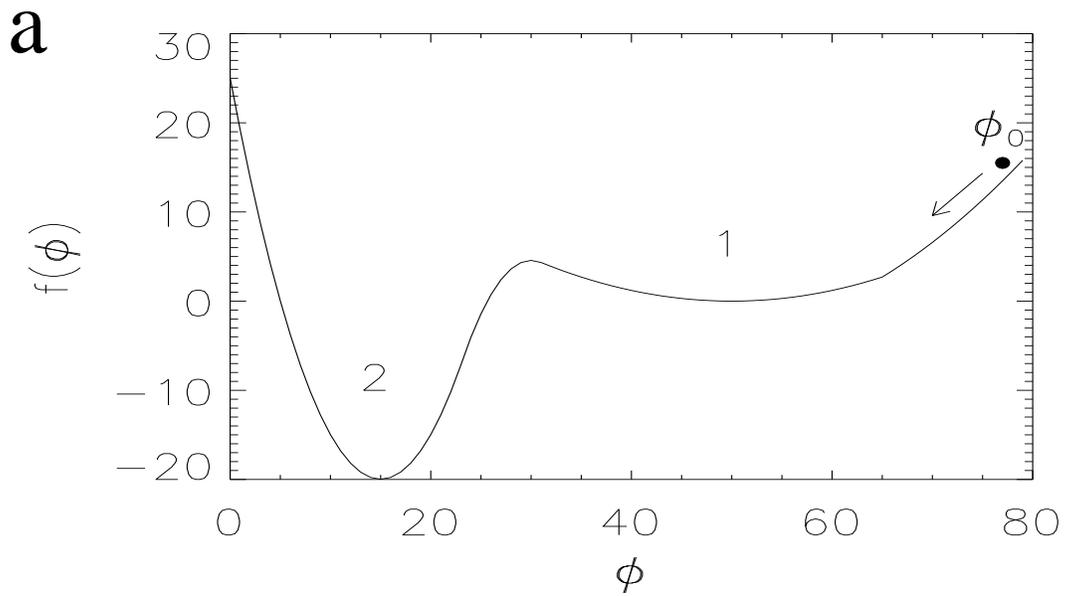
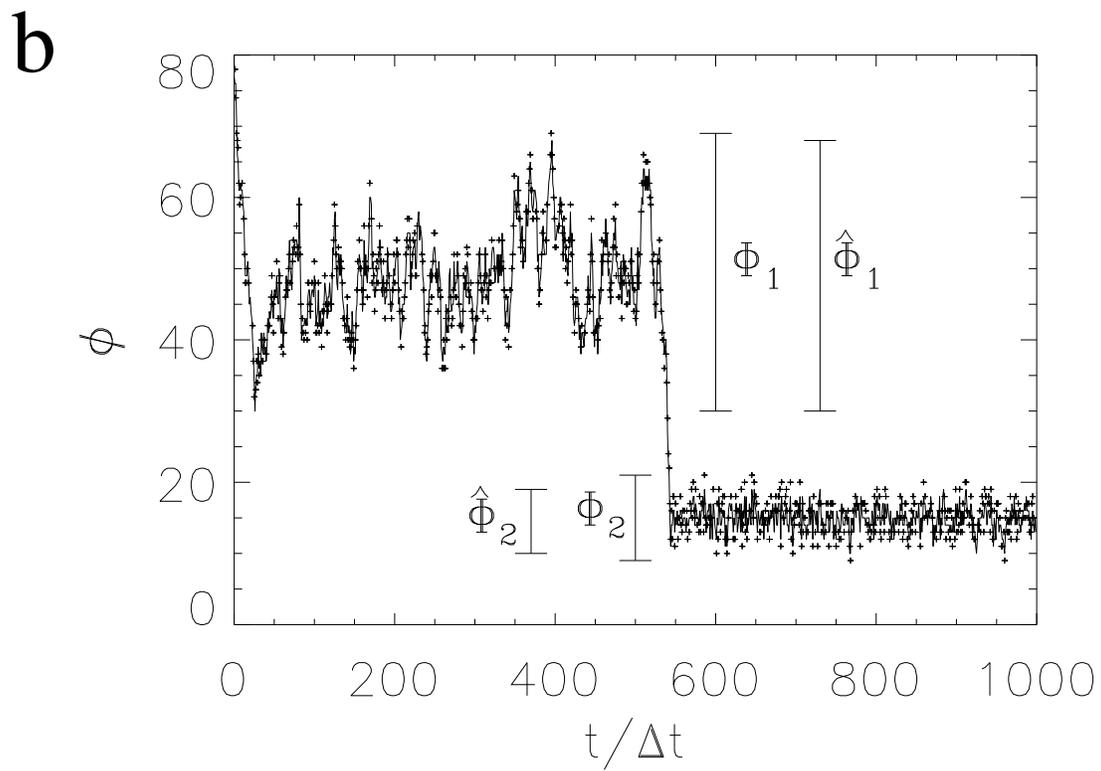
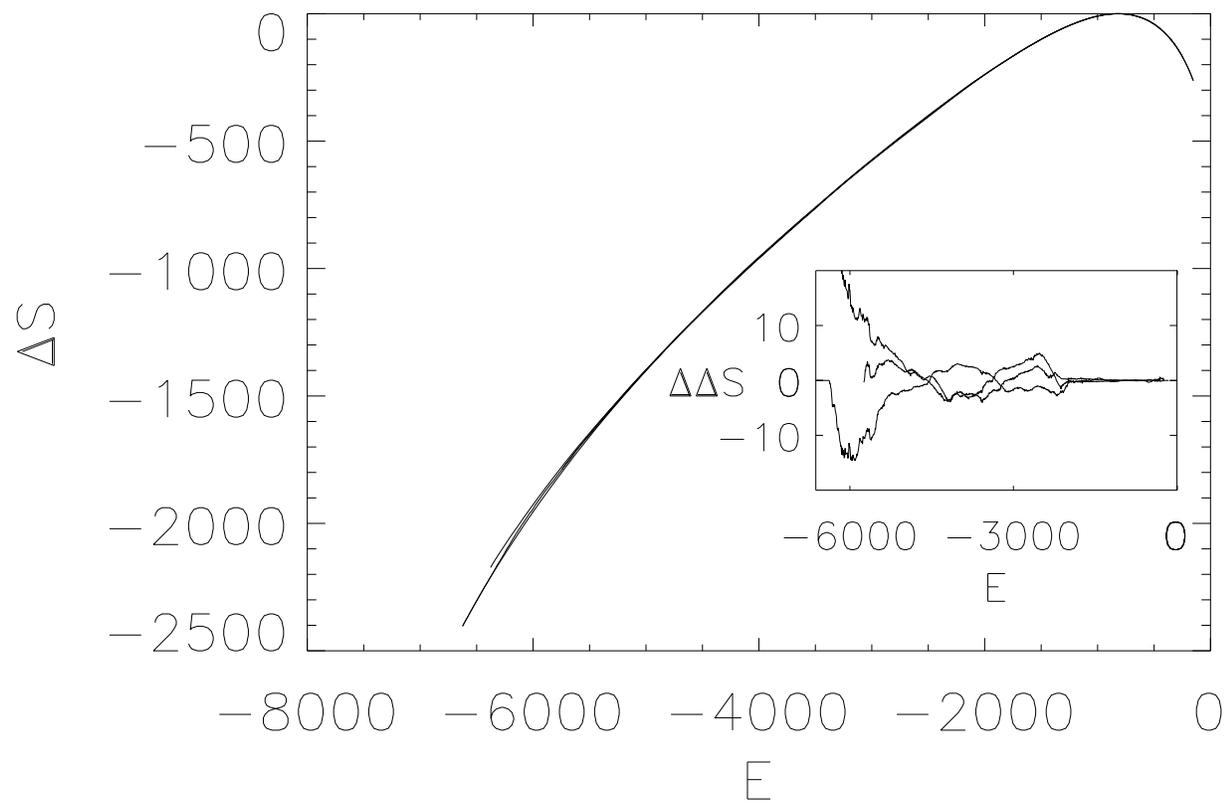
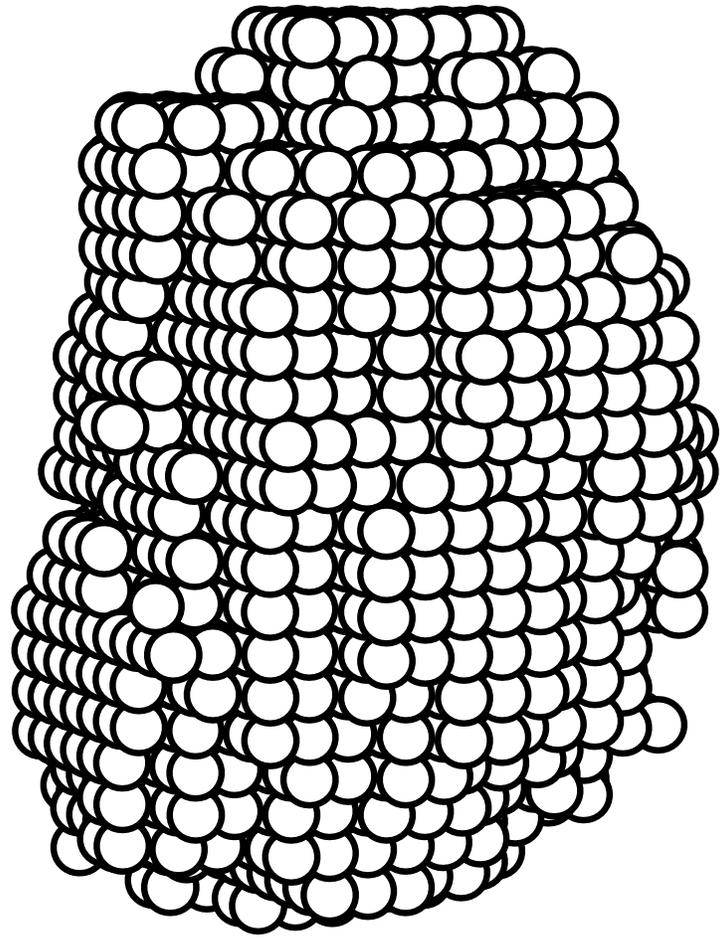
\end{document}